\documentclass[aps,prl,prev,twocolumn,superscriptaddress,floatfix,nofootinbib]{revtex4-1}
\pdfoutput=1

\usepackage{graphicx}
\usepackage{bm}
\usepackage{times}
\usepackage{slashed}
\usepackage{color}
\usepackage{aas_macros}
\usepackage{slashed}
\usepackage{lipsum}
\usepackage{subfigure}
\usepackage{multirow}
\usepackage{amsmath}
\usepackage{hyperref} 
\usepackage[normalem]{ulem}
\hypersetup{colorlinks = true, 
citecolor=blue,
linkcolor=blue,
filecolor=blue,      
urlcolor=blue
}

\begin{document}
\title{Origin of the Cosmic Ray Galactic Halo \\ Driven by Advected Turbulence and Self-Generated Waves}
\author{Carmelo Evoli}
\email{carmelo.evoli@gssi.it}
\affiliation{Gran Sasso Science Institute, Viale F.~Crispi 7, L'Aquila, Italy}
\affiliation{INFN/Laboratori Nazionali del Gran Sasso, Via G.~Acitelli 22, Assergi (AQ), Italy}

\author{Pasquale Blasi}
\email{pasquale.blasi@gssi.it}
\affiliation{Gran Sasso Science Institute, Viale F.~Crispi 7, L'Aquila, Italy}
\affiliation{INFN/Laboratori Nazionali del Gran Sasso, Via G.~Acitelli 22, Assergi (AQ), Italy}
\affiliation{INAF/Osservatorio Astrofico di Arcetri, L.go E.~Fermi 5, Firenze, Italy}

\author{Giovanni Morlino}
\email{giovanni.morlino@gssi.it}
\affiliation{Gran Sasso Science Institute, Viale F.~Crispi 7, L'Aquila, Italy}
\affiliation{INFN/Laboratori Nazionali del Gran Sasso, Via G.~Acitelli 22, Assergi (AQ), Italy}
\affiliation{INAF/Osservatorio Astrofico di Arcetri, L.go E.~Fermi 5, Firenze, Italy}

\author{Roberto Aloisio}
\email{roberto.aloisio@gssi.it}
\affiliation{Gran Sasso Science Institute, Viale F.~Crispi 7, L'Aquila, Italy}
\affiliation{INFN/Laboratori Nazionali del Gran Sasso, Via G.~Acitelli 22, Assergi (AQ), Italy}

\begin{abstract}
 The diffusive paradigm for the transport of Galactic cosmic rays is central to our understanding of the origin of these high energy particles. However it is worth recalling that the normalization, energy dependence and spatial extent of the diffusion coefficient in the interstellar medium are fitted to the data and typically are not derived from more basic principles. Here we discuss a scenario in which the diffusion properties of cosmic rays are derived from a combination of wave self-generation and advection from the Galactic disc, where the sources of cosmic rays are assumed to be located. {We  show for the first time that a halo naturally arises from these phenomena, with a size of a few kpc, compatible with the value that typically best fits observations in simple parametric approaches to cosmic ray diffusion. We also show that transport in such a halo results in a hardening in the spectra of primary cosmic rays at $\sim 300$ GV}.
\end{abstract}

\maketitle

{\it Introduction -- } Understanding cosmic-ray (CR) propagation in the Galaxy and its implications for observations at different energies and with different messengers is one of the challenges of modern astroparticle physics.

The standard scenario adopted to describe Galactic propagation in terms of properties of the interstellar turbulence is the so called {\it galactic halo model} proposed by Ginzburg and Syrovatskii in 1964~\citep{1964ocr..book.....G} and described in detail in~\citep{1990acr..book.....B}.
The halo model is usually implemented assuming that CRs are produced by sources located in the thin Galactic disc and then diffuse by scattering off random magnetic fluctuations in a low-density confinement region (``halo'') extending well beyond the gaseous disc. The size of this region is usually set by hand and chosen to fit observations.
Outside the magnetic halo, the turbulence level is assumed to vanish so that particles can escape freely into intergalactic space so as to reduce the CR density to $\sim 0$. 

From the theoretical point of view a problem of paramount importance is that of connecting the properties of the magnetic turbulence with particle diffusion, which in a generic turbulence (even an isotropic one) turns out to be anisotropic. 

Evidence of Kolmogorov-like turbulence across more than ten orders of magnitude in wave number is obtained from the observation of interstellar medium (ISM) scintillation~\citep{1995ApJ...443..209A} and of fluctuations of the Faraday rotation measurements~\citep{1996ApJ...458..194M}. The properties of magnetic turbulence that are relevant for particle diffusion are however not accessible to this type of observation and to date such turbulence and the corresponding diffusion properties of CRs remain poorly constrained. 
On the other hand, CR measurements allow one to define volume integrated properties of the turbulence through measurements of the boron to carbon (B/C) ratio and other secondary to primary ratios (see, e.g.~\cite{2010APh....34..274D}). These observations strongly suggest that CRs diffuse on a region of size $H$ that is much larger than the size of the Galactic gaseous disc, with half thickness $h$, in order to guarantee that the grammage traversed by CRs is large enough to produce the observed fluxes of secondaries. 

Recent precise measurements of the B/C ratio by AMS-02 can be accommodated at rigidities $R \gtrsim 60$~GV assuming a CR grammage that scales with rigidity as $R^{-1/3}$, that is claimed to be consistent with the diffusion coefficient expected from transport in a turbulence with Kolmogorov spectrum, $D(R) \sim 10^{28}(R/{\rm GV})^{1/3}$~cm$^2$~s$^{-1}$.

An independent piece of evidence of the existence of a magnetized halo comes from observations in the radio band of diffuse synchrotron emission, revealing the presence of electrons and magnetic fields above and below the Galactic plane~\cite{1985A&A...153...17B}. 
The existence of a halo of several kpc size can be inferred from a comparison between numerical models for the CR electron distribution and the morphology of the radio emission~\cite{2013MNRAS.436.2127O,2013JCAP...03..036D}. 
It is worth mentioning that radio halos with a similar size have been observed in other spiral galaxies~(e.g. NGC 4631, NGC 891).
In addition, the gamma-ray emissivity as a function of height above the disk $z$ can be inferred from gamma-ray observations of high-velocity clouds carried out by Fermi-LAT. The result reveals a confinement region for the CR hadronic component with size $H \gtrsim 2$~kpc, with a rather large uncertainty~\cite{2015ApJ...807..161T}. 

Existing measurements of both primary and secondary CR can be decently reproduced within the halo model, at least in the kinetic energy range $0.1 \lesssim T \lesssim 100$~GeV/n, although {\it ad hoc} breaks in either the injection spectra or the diffusion coefficient are needed to achieve a consistent picture. 

Additional {\it ad hoc} breaks are needed ~\cite{2016ApJ...824...16J} to accommodate the recent measurements of the proton and helium spectra recently carried out by PAMELA ~\cite{2011Sci...332...69A} and AMS-02~\cite{2015PhRvL.114q1103A,2015PhRvL.115u1101A}. 

While this is an effective approach to understanding some aspects of the origin of CRs, there is little doubt that it is highly unsatisfactory in terms of the basic physical aspects of the transport of CRs. First, breaks in otherwise power law trends typically signal the onset of new and potentially interesting physical phenomena. Second, in all current CR transport models, the size $H$ of the region where CRs are diffusively confined is an external parameter to be fixed to fit the grammage inferred from the observed flux of secondary stable  and unstable nuclei. Third, diffusion in the ISM is usually treated in a simplified way, so that particles diffuse isotropically in all directions or just in the direction perpendicular to the Galactic disc (1D models). 

In the following we address the first two issues listed above, proposing possible ways to gain insights into the origin of CR diffusion and aiming at achieving a physical understanding of how the CR halo might arise. 
The breaks in the spectra of primary elements are most easily understood as a consequence of intervening phenomena in CR transport rather than associated with either acceleration or random effects in the spatial distribution of the sources \cite{2017PhRvL.119x1101G}. 

In \cite{2012ApJ...752L..13T,2015PhRvD..92h1301T}, the hardening in the spectra of protons and helium has been attributed to a spatial dependence of the diffusion coefficient: Two regions in an otherwise fixed halo of size $H$ are assumed to exist, and the diffusion coefficient in the two regions is chosen so as to fit the observed spectra.
On the other hand, in \cite{2012PhRvL.109f1101B,2013JCAP...07..001A,2015A&A...583A..95A}, the spectral breaks were explained as a consequence of a transition from self-generated to pre-existing turbulence. In these nonlinear approaches to CR transport the diffusion coefficient is an output of the calculations, as derived from the CR gradients that are responsible for the excitation of streaming instability. 

However, also in these approaches the halo size is fixed and the possible spatial dependence of the diffusion coefficient in the halo is not accounted for. It is likely that both these phenomena are at work at the same time. 

Here, we propose a physical view of how CR diffusion occurs in a halo that arises naturally rather being imposed by hand: the waves that CRs scatter off are considered as self-generated by the same CRs, according to the local gradient, and advected outward at the local Alfv\'en speed. At the same time, sources in the Galactic disc (for instance supernova explosions) are also assumed to inject turbulence on large scales ($\sim 10-100$ pc). Such turbulence is then advected away from the disc and at the same time cascades towards smaller scales. Both effects induce a spatial dependence in the diffusion coefficient. 
We describe the cascading as a nonlinear diffusion process in $k$-space, as proposed in \cite{1995ApJ...452..912M}, although this approach does not include anisotropic cascading that is known to occur~\cite{2002ApJ...564..291C}. On the other hand, such anisotropy is seen to become prominent after the turbulence has cascaded down to values of $k$ larger than the injection scale $k_0$ by 1-2 orders of magnitude. This corresponds, for typical parameters of resonant scattering in the Galaxy, to particles with energy below $\sim 10$ TeV, where the transport starts to be affected by the self-generation process induced by streaming instability.

Some previous attempts to investigate the nature of the halo \cite{1993A&A...268..356D,1994A&A...281..937D} were made by assuming spherical symmetry and trying to describe the halo boundary as the location where anisotropy becomes of order unity, expected in $f\to 0$ at $|z|>H$, mimicking free escape. As we discuss below this scenario is quite different from the one we find here, where the diffusion coefficient increases with $z$ as a power law, so that no definite boundary exists where $f\to 0$.

In order to describe the nonlinear chain of phenomena introduced above we solve numerically two coupled time-dependent nonlinear partial differential equations, one describing the transport of CRs and the other describing the excitation, advection, and cascading of waves. 

{\it CR transport in self-generated diffusion --}
The transport of CRs is well described by the advection-diffusion equation:
\begin{multline}\label{eq:transport}
\frac{\partial f}{\partial t} 
-\frac{\partial}{\partial z}\left( D_{zz} \frac{\partial f}{\partial z}\right) 
+ v_{\rm A}\frac{\partial f}{\partial z} 
- \frac{dv_{\rm A}}{dz}\frac{p}{3} \frac{\partial f}{\partial p}
\\ + \frac{1}{p^2}\frac{\partial}{\partial p} \left[ p^2 \left( \frac{dp}{dt} \right)_{\rm ion} f \right]
= Q_{\rm CR},
\end{multline}
where $f(p,z,t)$ is the phase space distribution function, $v_{\rm A}$ is the Alfv\'en speed, and $\dot{p}_{\rm ion}$ is the rate of ionization losses, typically important for low energy CRs. 
As in previous calculations, e.g., \cite{2012PhRvL.109f1101B,2013JCAP...07..001A}, we make the simplifying assumption that transport occurs only in the $z$ direction.
CR injection is assumed to take place inside a disk of radius $R_d = 10$~kpc and with a Gaussian profile along $z$ with the same width as the gas disk, $\sigma = 100$~pc.

Having in mind SNRs as the sources of CRs, the source term is normalized to the surface density rate of SN explosions, while the injection spectrum $\Phi(p)$ is assumed to be in the form of a power law in momentum with the same slope $\alpha$ at any location inside the disc.
The normalization of the spectrum is chosen so as to have a fraction $\xi_{\rm CR}$ of the total energy of a supernova $E_{\rm SN}$ channeled to CRs (see also \cite{2016MNRAS.462.4227R}). 

The diffusion coefficient that appears in Eq.~\ref{eq:transport} can be obtained using quasilinear theory:
$D_{\rm zz}(z,p) = \frac{\beta(p) c r_L(p)}{3}\left[ \frac{1}{k_{\rm res}W(k_{\rm res})} \right]$ ,
where $W(k_{\rm res})$ is the spectrum of turbulence calculated at the resonance wave number $k_{\rm res} = 1 / r_L(p)$ and normalized to the regular field energy density, $U_{B_0} = B_0^2 / 8 \pi$. 

The third and the fourth terms in Eq.~\ref{eq:transport} describe the advection of CRs with waves propagating with velocity $v_A$. The implicit assumption here is that waves are all moving away from the disc, in the $z$ direction. 

We consider two sources of waves responsible for CR scattering, namely, (1) waves deriving from turbulent cascading of power injected by SNRs at large scales and (2) self-generated waves produced by CRs through streaming instability.
The transport equation for the Alfv\'en wave spectral energy density is~\cite{1993ApJ...413..619J} 
\begin{equation}
\frac{\partial W}{\partial t} 
+ \frac{\partial}{\partial k} \left( D_{kk} \frac{\partial W}{\partial k} \right) 
+ {\frac{\partial}{\partial z} \left( v_A W \right)} 
= {\Gamma_{\rm CR} W} + \frac{Q_W(k,z)}{U_{B_0}}.
\label{eq:waves}
\end{equation}

The turbulent cascading is described in terms of diffusion in $k$-space with diffusion coefficient $D_{kk} = c_k |v_A| k^{7/2} W^{1/2}$, where $c_k \sim 5.2 \times 10^{-2}$ is a constant~\cite{1995ApJ...452..912M}.
Notice that this diffusion is nonlinear in that the diffusion coefficient depends on the power spectrum $W(k)$. 

The third term in Eq.~\ref{eq:waves} represents an advective transport of waves along $z$, with a speed that equals the Alfv\'en speed $v_A$. A note of caution is in order: formally, the advection velocity coincides with $v_A$ only for waves that are self-generated, since in this case they are all produced in the same direction. Waves deriving from turbulent cascading move in both directions along the $z$ axis. On the other hand the symmetry of the problem (sources all located around $z\sim 0$) suggests that the net advection velocity of such waves away from the disc is somewhat smaller than $v_A$ but still close to it. Lacking a better description of advection, we retain $v_A$ as the net velocity of all waves in our problem. 

\begin{figure}
\centering
\includegraphics[width=0.47\textwidth]{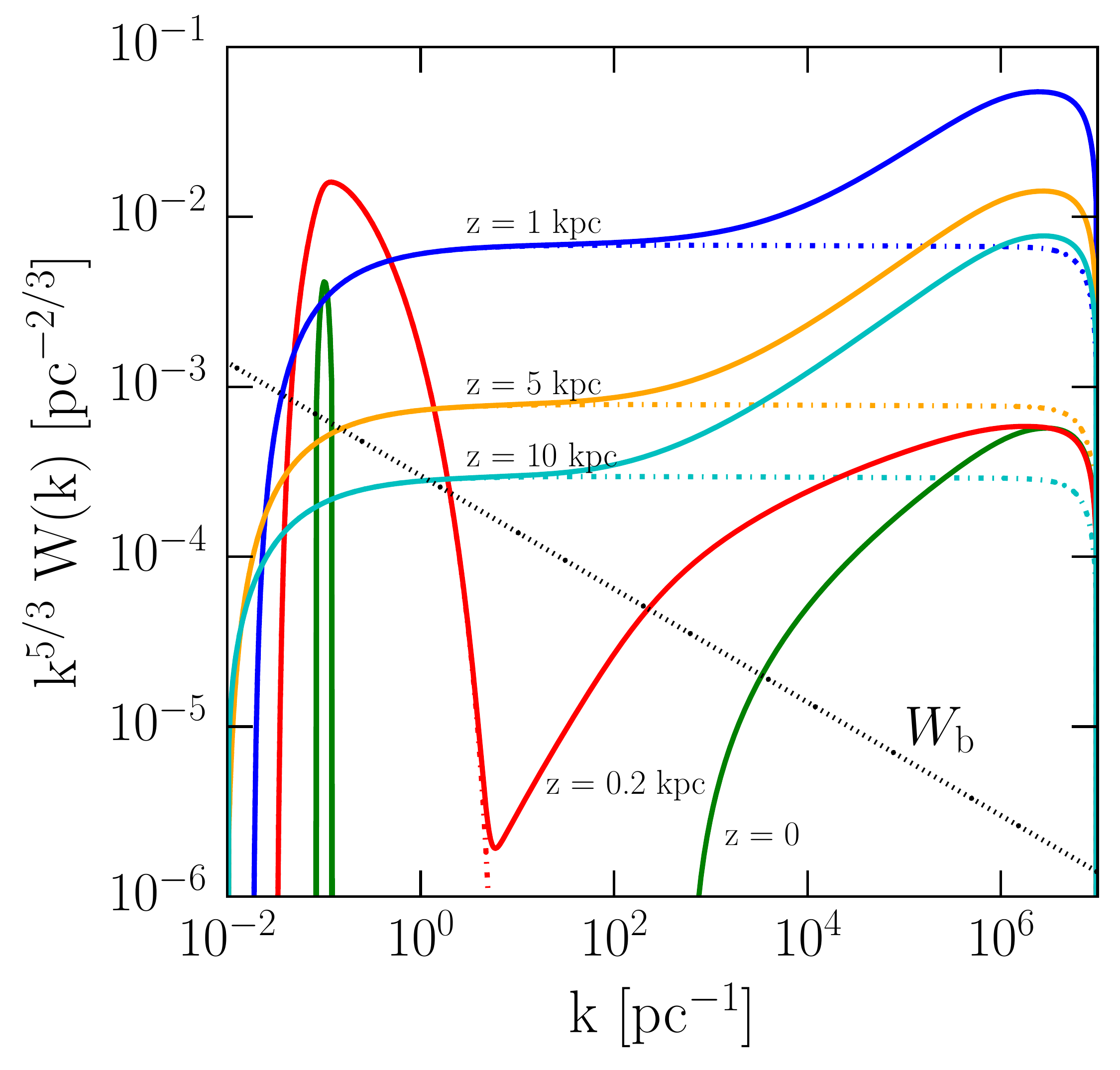}
\caption{Wave spectrum as a function of $k$ with (solid lines) and without (dashed lines) the contribution of self-generated waves at different positions. At z = 0 solid and dashed lines overlap around the injection peak at $k_0$. 
The dotted line shows the background turbulence we add to constrain diffusion velocity to be smaller than $c$.}
\label{Fig:W_spectrum_0}
\end{figure}

Waves may be generated by CRs through streaming instability, with a rate:
$\Gamma_{\rm CR}(k) = \frac{16\pi^2 v_A}{3kW(k)B_0^2} \left[p^4 v(p) \left| \frac{\partial f}{\partial z} \right| \right]_{p_{\rm res}}$ 
that reflects the dependence of the wave growth on the CR gradient $\frac{\partial f}{\partial z}$. Here $p_{res}(k)$ is the momentum of particles that can resonate with waves with wave number $k$~\citep{1971ApJ...170..265S}.

Finally, we introduce in Eq.~\ref{eq:waves} an injection term that should mimic the generation of waves by, for instance, supernova explosions:
$
Q_W(z,k) = Q_0 \xi_{\rm W} \delta(k - k_0) \frac{{\rm e}^{-z^2/\sigma^2}}{\sqrt{2\pi\sigma^2}},
$
where the injection scale is $k_0 = 0.1~{\rm pc}^{-1}$ (corresponding to $l_0 = 10~{\rm pc}$) and the conversion efficiency is $\xi_{\rm W} = 10^{-4}$.

The equations for the waves and for CR transport are solved in an iterative way. 
The procedure is repeated until convergence is reached, which typically requires a few steps. 

\begin{figure}
\centering
\includegraphics[width=0.47\textwidth]{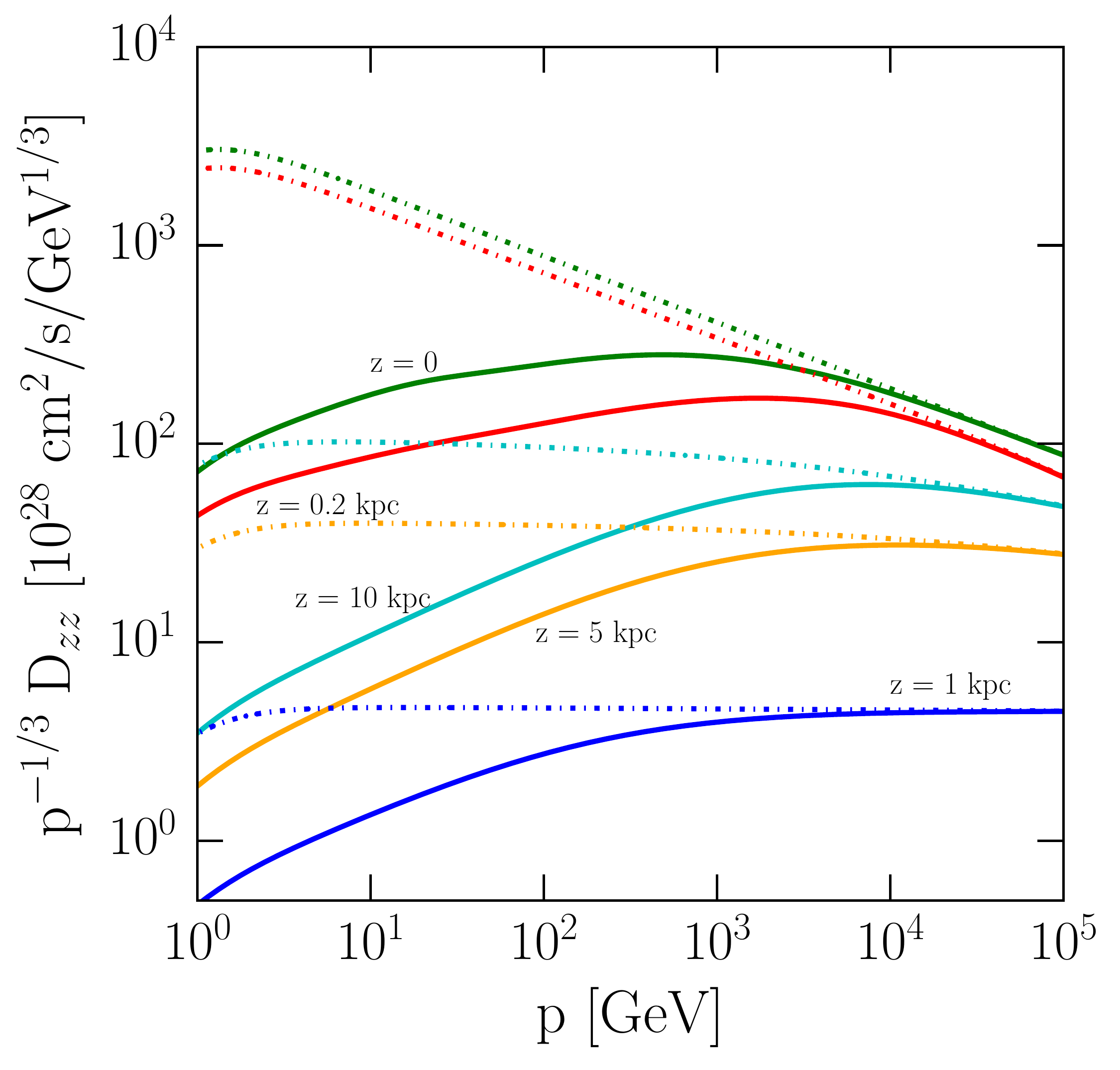}
\caption{Diffusion coefficient as a function of momentum for different values of $z$.}
\label{Fig:diffusion_halo}
\end{figure}

{\it Results --} 
We start the discussion of our results from the case with no self-generation, namely a case in which waves are only injected at some scale $k_0^{-1}$ in the disc of the Galaxy and cascade in wave number space with a Kolmogorov phenomenology, while advecting away from the disc with Alfv\'en speed. This case helps us illustrate the phenomenon of self-determination of a halo, as due to the cascading of power from the largest scale to the ones relevant for CR diffusion. 
Numerically, we seek the steady-state solution of Eq.~\ref{eq:waves} assuming $\Gamma_{\rm CR} = 0$ on a spatial grid $0\leq |z|\leq 20$ kpc. We check {\it a posteriori} what is the effect of changing the size of the box in which the solution is found, in order to make sure that our results do not reflect a numerical boundary condition.

The spectrum of waves is shown in Fig.~\ref{Fig:W_spectrum_0} for different locations away from the disc. One can see that within a few hundred pc from the Galactic disc the power remains concentrated around the injection scale, while at larger distances the nonlinear cascading populates the large-$k$ region of the spectrum, with a slope that is very close to $5/3$, typical of a Kolmogorov spectrum. On kpc scales, the slope of the spectrum remains stable, while the normalization drops, as a result of the cascade that transports power down to the dissipation scale. 
It is important to keep in mind that since the cascade is nonlinear the quantitative details depend on the specific realization of the problem. For instance, increasing the rate of injection leads to an increase of $W(k_0)$ and hence to a shorter characteristic time for the cascading process, $\tau_c\sim k_0^2/D_{kk}(k_0)$. The region $k>k_0$ is populated only at $z\gtrsim v_A \tau_c \sim$ several kpc's.

In order to avoid numerical problems that occur due to regions where $W(k)$ is vanishingly small, we impose a physical constraint: the diffusion coefficient cannot be larger than the one that corresponds to motion at the speed of light on a region of size $H$, namely $D_H=(1/3) c H$. This can be considered as due to a fictitious power spectrum $W_b$ that is shown in Fig.~\ref{Fig:W_spectrum_0} as a dotted line. This trick turns out to be necessary only very close to the disc where turbulence on small scales (resonant with CRs with the energies we are interested in) do not have time to develop through cascading.
The relevant diffusion coefficient as a function of particle momentum is shown in Fig.~\ref{Fig:diffusion_halo} as dashed lines. 

\begin{figure}
\centering
\includegraphics[width=0.47\textwidth]{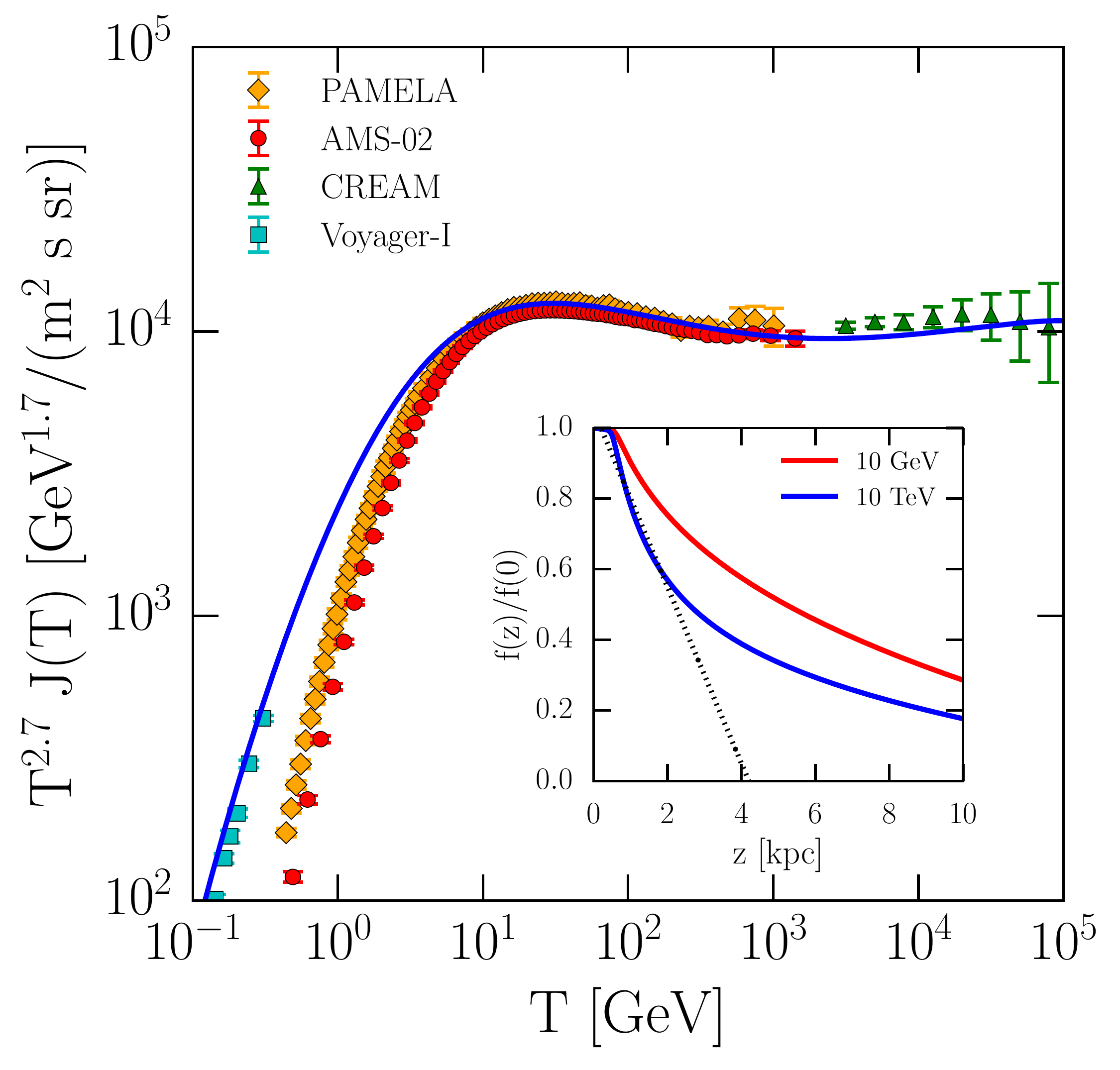}
\caption{Spectrum of protons in the local ISM compared to observational data. The spatial dependence of the CR distribution function is shown in the inset, for energies 10 GeV and 10 TeV.}
\label{Fig:f_spectrum}
\end{figure}

At this point we introduce the contribution of self-generated waves, as due to CR streaming. This phenomenon adds to the nonlinearity of the problem, in that the amount of waves produced by this phenomenon is related to the number density and gradients of CRs, which are in turn the result of the scattering of CRs on self-generated (or preexisting) waves. The rate of self-generation and the rate of CR injection by sources in the Galaxy are clearly related to each other and need be calibrated to the observed spectrum of CRs. 

It is worth noticing that in the near-disc regions, where the nonlinear cascade has no time to develop down to the small scales resonant with CR energies, CR scattering is fully determined by self-generated waves. In the distant regions, depending on particle energy, self-generation and nonlinear cascading compete with each other. This competition results in breaks in the spectra of primary CRs, as a result of both a complex power spectrum of the turbulence and of the spatial dependence of the diffusion coefficient. These effects are illustrated in Fig.~\ref{Fig:W_spectrum_0} (solid lines), where one can see that in the presence of self-generation the power spectrum is enriched with power in the high $k$ range, with respect to the simple cascade from larger spatial scales. In the near disc region ($|z| \lesssim 0.2$~kpc), virtually all the power at the resonant scales with CRs in the energy range below $\sim$TeV is due to self-generation. 

In terms of CR transport, these effects are more clearly visible in Fig.~\ref{Fig:diffusion_halo}: only at very high energies CR scattering is due to Kolmogorov turbulence (solid and dashed lines overlap), while at basically all distances from the disc, scattering is mainly due to self-generated waves for $E\leq 1$ TeV. 
 
The spectrum of protons as calculated solving the set of equations describing CR and wave transport together is plotted in Fig.~\ref{Fig:f_spectrum}, as compared with data from PAMELA~\cite{2011Sci...332...69A}, AMS-02~\cite{2015PhRvL.114q1103A} and CREAM~\cite{2011ApJ...728..122Y} at high energies and Voyager~1 \cite{2013Sci...341..150S} at low energies. 

The inset in the same figure shows the spatial dependence of the solution for two values of energy (10 GeV and 10 TeV), compared with the linear decrease predicted in the standard halo model with a halo size $H=4$ kpc.

In the range of energies $10 \lesssim T \lesssim 200$ ~GeV/n the self-generation is so effective as to make the diffusion coefficient have a steep energy dependence.
As a consequence the injection spectrum that is needed to fit the data is $p^2 \, \Phi(p) \propto p^{-2.2}$, which is not far from what can be accounted for in terms of DSA if the velocity of the scattering centers is taken into account \cite{2012JCAP...07..038C}.
At lower energies the CR transport becomes dominated by advection with Alfv\'en waves. In this regime advection and ionization losses make the spectrum in the disc close to the injection spectrum. 

The CR acceleration efficiency in terms of protons that is needed to ensure the level of wave excitation necessary to explain observations, is $\epsilon_{\rm CR} \sim 4$\%, in line with the standard expectation of the so-called SNR paradigm. 

{\it Conclusions -- } {We use a numerical approach to the solution of the transport equations for particles and waves to show that the CR halo arises naturally from a combination of the turbulence injected in the Galactic disc and eventually advected into the halo, and self-generated waves due to the excitation of streaming instability through CR gradients. This finding addresses the long standing issue that in the context of the traditional halo model the CR spectrum observed at the Earth reflects the free escape boundary condition at the edge of the halo, imposed by hand.}

The turbulent cascade introduces a scale $z_c \approx v_A k_0^2/D_{kk}$ below which turbulence is mainly self-generated. At larger distances the cascade quickly develops and leads to a rough space dependence of the diffusion coefficient $\propto z^{\alpha}$ with $\alpha \gtrsim 1$. As a consequence, the spectrum in the disc depends on the scale $z_c$ but only weakly on the artificial boundary at $z=\pm H\gg z_c$. Moreover, for typical values of the parameters, one has $z_c \sim$ few kpc, and $z_c$ plays the role of an effective size of the halo. The observed spectral break at rigidity $\sim 300$ GV also arises naturally, because of a transition from a diffusion dominated by self-generation (at lower energies) to a Kolmogorov-like diffusion at higher energies.

As noticed in~\cite{2013JCAP...07..001A}, chemicals heavier than protons can contribute to self-generation (helium nuclei provide a contribution similar to that of protons, while heavier nuclei account for about 10\% of self-generated waves) and will be included in future generalizations of this work.

Both the cascade and the self-generation of waves by CRs are nonlinear processes: the combination of the two leads to an interpretation of the observed halo as a by-product of a self-regulation process that is typical of nonlinear phenomena. 

The authors are very grateful to Elena Amato for numerous discussions on the topics of the present paper. C.E.~acknowledges the European Commission for support under the H2020-MSCA-IF-2016 action, Grant No.~751311 GRAPES – Galactic cosmic RAy Propagation: an Extensive Study.

\bibliography{selfgeneration}
\end{document}